\documentclass[twocolumn,showpacs,superscriptaddress,eqsecnum,twoside,pra]{revtex4}
\usepackage{mathrsfs}
\usepackage{amssymb, amsbsy, amsmath, latexsym, dsfont, array, layout, graphicx,mathrsfs,color}

\newcommand{\one}{\mathds{1}}
\newcommand{\ket}[1]{\left|{#1}\right\rangle}
\newcommand{\bra}[1]{\left\langle{#1}\right|}

\newcommand{\dslash}{\not{\text{d}}}

\begin{document}

\title{Controlling and reversing the transition from classical diffusive\\
    to quantum ballistic transport in a quantum walk by driving the coin}
\author{Peng Xue}
\affiliation{Department of Physics, Southeast University, Nanjing
211189, China}
\author{Barry C. Sanders}
\affiliation{Institute for Quantum Science \& Technology, University of Calgary, AB T2N 1N4, Canada}
\date{\today}

\begin{abstract}
We show that the standard quantum-walk quantum-to-classical transition,
characterized by ballistic-to-diffusive spreading of the walker's position,
can be controlled by externally modulating the coin state.
We illustrate by showing an oscillation between classical diffusive and quantum ballistic
spreading using numerical and asymptotically exact
closed-form solutions,
and we prove that the walker is in a controllable incoherent
mixture of classical and quantum walks  with a reversible quantum-to-classical transition.
\end{abstract}


\maketitle

\section{Introduction}
\label{sec:intro}

The quantum walk on a lattice~\cite{Kem03a,Kem03b} is an important
branch of quantum information research for several
reasons~\cite{Ken11}. The quantum walk concept drives breakthroughs
in quantum algorithms, including speeds up for quantum
searches~\cite{San08} and exponential speed-ups of graph traversal
compared with the best known classical algorithms for such
tasks~\cite{CCD+03,SKW03}. Beyond quantum algorithms and into the
physical world, quantum walks are evident in spin-chain quantum
transport~\cite{Kay10} and photosynthetic excitonic energy
transport~\cite{MRLA08}. Quantum walks serve as one model for
quantum computation~\cite{Chi09,CGZ12} alongside other models such
as circuit~\cite{Deu85}, measurement-based~\cite{RB01},
adiabatic~\cite{FGG+01} and topological quantum
computing~\cite{NSS+08}. Experimental realizations of quantum walks
abound with successes having been reported in nuclear magnetic
resonance~\cite{RLBL05}, ion traps~\cite{ZKG+10} and
photons~\cite{BFL+10,SCP+11}.

Decoherence is especially important in quantum
walk implementations, both because it deleteriously destroys unitarity with
consequences such as transforming the ballistic spreading to
diffusive spreading~\cite{BCA03a,BCA03b,KS05} and because of the
beneficial property of enabling tuning of quantum walk
dynamics~\cite{Ken07}.
Typically decoherence is characterized by the
rate of spreading of the walker's position after tracing over the
coin state, and ballistic spreading is `quantum' and diffusive
spreading is `classical'.

Here we show that controlling the walker's
coin in an appropriate time-dependent way enables the walker to
achieve diffusive spreading and later back to ballistic, i.e.\
transferring between classical and quantum behavior in a controlled
way.
Our result is distinct from studies of recurrences in quantum-walk dynamics~\cite{SJK08}
and from unitarily controlled ``stroboscopic'' quantum walks~\cite{BB04} or, equivalently,
modified quantum walk dynamics by periodic perturbations~\cite{WLK+04}.
Those investigations are essentially about unitary control.

Our result,
based on controlled non-unitary evolution through coin measurements,
instead challenges the paradigm of the irreversibility of the ballistic-to-diffusive spreading rate due to coin measurement.
On the other hand we show that the sacrosanct principle of non-decreasing walker-distribution entropy remains intact.

\section{Formalism}
\label{sec:formalism}

The joint walker-coin state~$\rho_\text{wc}$ is a trace-class
positive operator on the Hilbert space
$\mathscr{H}=\mathscr{H}_\text{w}\otimes\mathscr{H}_\text{c}$ such
that $\mathscr{H}_\text{w}=\text{span}\{|x\rangle;x\in\mathbb{Z}\}$
with $|x\rangle$ the orthonormal walker position states on a regular
integer lattice, and
$\mathscr{H}_\text{c}=\text{span}\{|\pm\rangle\}$ with $|\pm\rangle$
the two coin states. If the walker-coin system undergoes periodic
unitary steps, the evolution operator is given by
\begin{equation}
    U:=F(\one\otimes C)
\end{equation}
for $\one$ the identity, $H$ the Hadamard operator,
\begin{equation}
    C:=H=\frac{1}{\sqrt{2}}\begin{pmatrix}1&1\\1 &-1\end{pmatrix},
\end{equation}
and
\begin{equation}
    F:=S\otimes\ket{+}\bra{+}+S^\dagger\otimes\ket{-}\bra{-}
\end{equation}
the conditional shift operator with shift operator
\begin{equation}
    S:=\sum_x\ket{x+1}\bra{x}.
\end{equation}
The walker's evolving state is
\begin{equation}
    \rho_\text{wc}(t)=\mathcal{U}^t\rho_\text{wc}(0):=(U^\dagger)^t\rho_\text{wc}(0)U^t
\end{equation}
for discrete time parameter~$t\in\mathbb{Z}$.
The driven-coin case is more complicated and treated below.

We now generalize for the case of the driven coin. Prior to each
unitary coin `flip', a completely-positive trace-preserving map
$\mathcal{E}(t,0):\rho_\text{c}(0)\mapsto\rho_\text{c}(t)$ is
applied to the coin. The map can be decomposed into the operator sum
\begin{equation}
    \mathcal{E}\rho_\text{c}=\sum_{n\in\{0,\pm\}}A_n\rho_\text{c} A^\dagger_n
\end{equation}
such that
\begin{equation}
    \sum_{n\in\{0,\pm\}}A^\dagger_n A_n=\one
\end{equation}
with
\begin{equation}
    A_0=\sqrt{\kappa(t)}\one,\;
    A_\pm=\sqrt{1-\kappa(t)}\mathcal{P}^\pm,
\end{equation}
and
\begin{equation}
    \mathcal{P}^\pm=\ket{\pm}\bra{\pm},
\end{equation}
emerge corresponding to a coin with probability $1-\kappa$ of being
measured at each step.
In other words,
$\kappa(t)$ is effectively a time-varying strength of coin-state measurement.

To understand the time-varying strength of the measurement,
let us consider a distinct but easy-to-undersand `off-on-off' model,
where $\kappa=1$ (no measurement hence perfect quantum walk) for some time,
then zero (strongest possible measurement, which reduces the dynamics to a classical random walk), then unity
(i.e., back to the quantum walk) again.
When the coin is unmeasured during the `off' interval, no dephasing takes place;
hence the quantum walk is unitary and the walker-coin state is an entangled pure state.

Now consider the second interval of this `off-on-off' model.
During this time the entangled pure state is subject to a strong coin-state measurement in the $|\pm\rangle$ basis.
Our unconditional measurement erases the knowledge of whether $+$ or $-$ were read
and instead projects the entangled state into a mixture of two product states of the walker
with the coin state: $|\phi_\pm\rangle|\pm\rangle$.
This state corresponds to whether the walker's motion is in the $+$ or $-$ direction in the last step~\cite{KS05}.

If the measurement were now turned off permanently,
each of these
states would evolve according to the quantum walk. Although mixing
would appear to erase some of the finer interference fringes, both
pure states would result in ballistic spreading of the walker. If
the strong measurement is on for two time steps, four pure states
$|\phi_{+\pm}\rangle|\pm\rangle$ and
$|\phi_{-\pm}\rangle|\pm\rangle$ corresponding to the four
measurement outcomes. Our unconditional measurement model loses
knowledge of the two coin measurements and thereby yields a mixture
of these four states and a narrow spread than for each walker
distribution individually because of regression to the classical
walk. Furthermore these states then evolve to produce ballistic
spreading of the walker's position.

We have considered just two time steps of strong measurement.
Now consider extrapolating to many times steps
corresponding to many sequential strong
measurements.
The consequence of this measurement taking place over~$t$ times steps
is exponentially many (in terms of~$t$) pure states all blended
together in one mixed state.
This mixed state converges to a Gaussian walker whose position spread reaches
classical-walk width.
Then, when the strong measurement ceases after~$t$ steps,
ballistic spreading recommences
for each element of the mixture hence the walker's distribution as a
whole.

This `off-on-off' model shows how sequential strong and weak measurements affect the dynamics
at an intuitive level, but having the walker position distribution alternately
increase and decrease over time requires a more sophisticated model.
As we show in the next section, the `periodically varying measurement strength' model
delivers such a result.

\section{Periodically varying measurement strength}

In this section, we introduce the `periodically varying measurement strength' model and
begin by analyzing numerically the effect of periodically varying the coin measurement strength.
Specifically we are concerned with the dynamically changing reduced walker position distribution,
and the time-dependent variance and entropy associated with this distribution.
For the periodic dynamical map with
\begin{equation}
    \kappa(t)=\cos\eta t,
\end{equation}
the corresponding state mapping is
\begin{equation}
 \mathcal{E}\rho_\text{c}
    =\kappa(t)\rho_\text{c}
        +[1-\kappa(t)]\left[\mathcal{P}^+\rho_\text{c}\mathcal{P}^++\mathcal{P}^-\rho_\text{c}
    \mathcal{P}^-\right]
\end{equation}
and
\begin{align}
 \rho_\text{c}^{00}\mapsto&\rho_\text{c}^{00},
    \rho_\text{c}^{01}\mapsto\cos\eta t\rho_\text{c}^{01},\nonumber\\
    \rho_\text{c}^{10}\mapsto&\cos\eta t\rho_\text{c}^{10},
    \rho_\text{c}^{11}\mapsto\rho_\text{c}^{11}.\nonumber
\end{align}
The walker's reduced state and resultant position distribution at time~$t$ are
\begin{equation}
    \rho_\text{w}(t)=\text{Tr}_\text{c}\rho_\text{wc}(t)
\end{equation}
with
\begin{equation}
    P_\text{w}(x,t)=\langle x|\rho_\text{w}(t)|x\rangle
\end{equation}
respectively.

If the coin state is initially
\begin{equation}
    \frac{\ket{+}+i\ket{-}}{\sqrt{2}}
\end{equation}
and $P_\text{w}(x,0)=\delta_{x0}$,
then $P_\text{w}(x,t)=P_\text{w}(-x,t)$. The walker's spread is
\begin{equation}
    \sigma(t):=\sqrt{V(t)},\;V=\langle x^2\rangle-\langle x\rangle^2
\end{equation}
and $\langle x^m\rangle(t)$ the $m^\text{th}$ moment
of~$P_\text{w}(x,t)$. Rate of spreading~$\sigma(t)$ is widely used
to differentiate between quantum and classical random walks. In
diffusive transport, $V \propto  t$, whereas $V\propto t^2$ for
ballistic transport, which holds for the coherent quantum
walk~\cite{AVWW11,AAM+12}.

Driving the coin with periodic~$\kappa(t)$
affects variance~$V$ as shown in Fig~\ref{fig:varanddist}(a):
\begin{figure}
    \includegraphics[width=0.45\columnwidth]{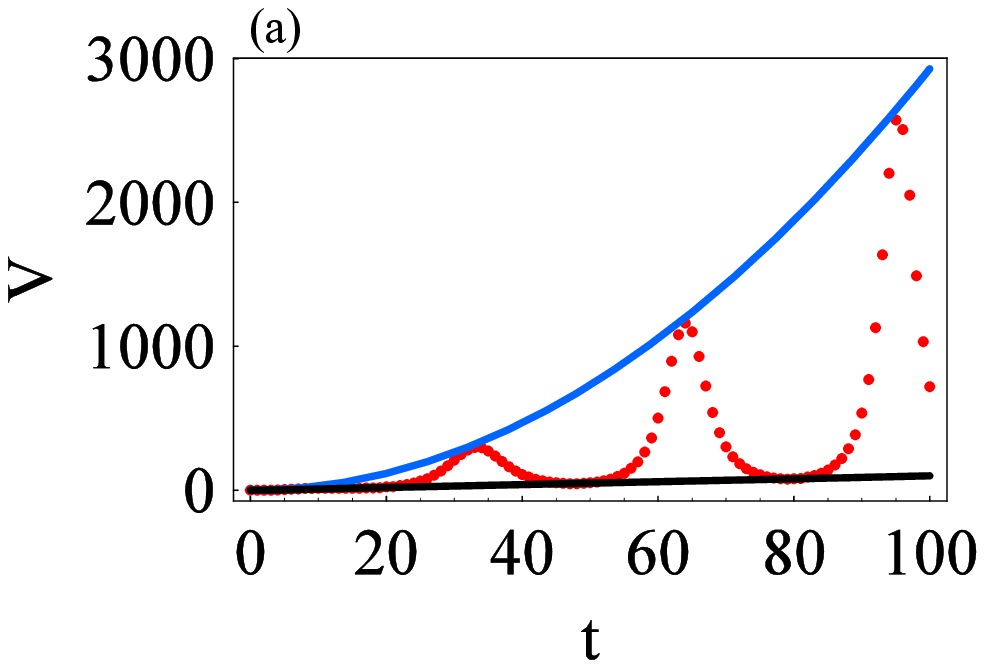}
    \includegraphics[width=0.45\columnwidth]{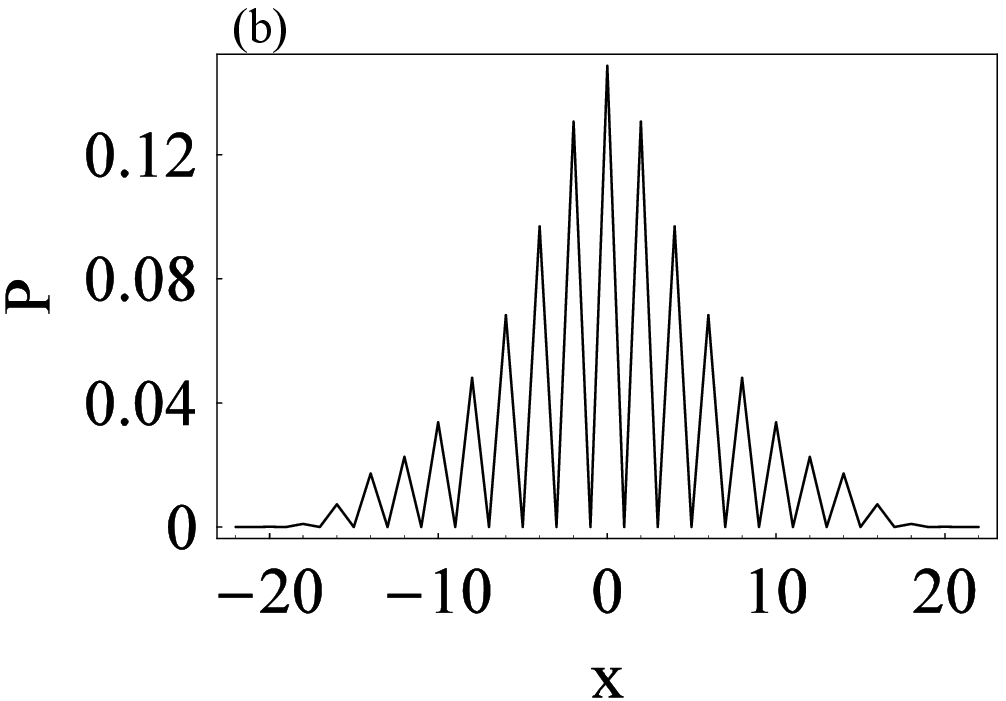}
    \includegraphics[width=0.45\columnwidth]{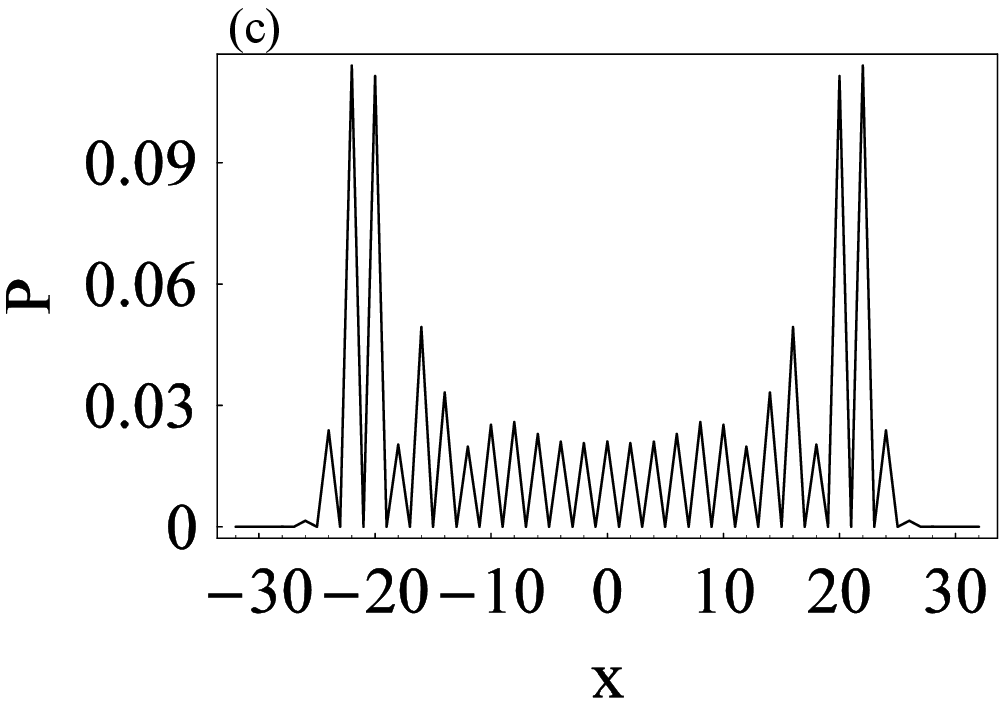}
    \includegraphics[width=0.45\columnwidth]{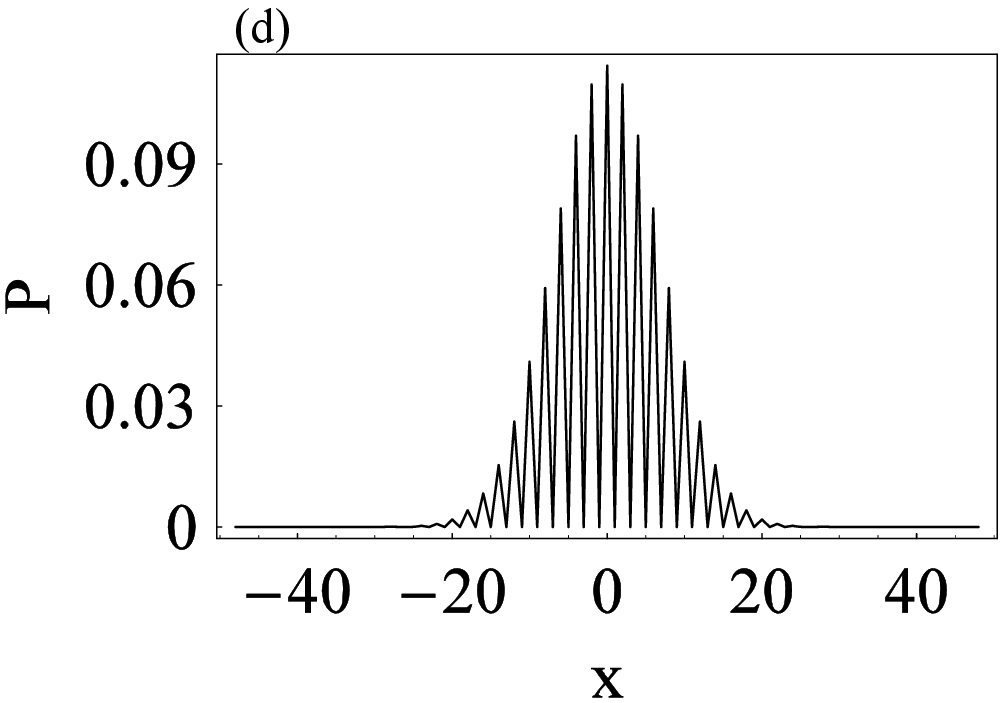} \caption{
        (Color online)
        (a)~Numerically evaluated variance of the walker's position over time~$t$
        with initial state $\ket{0}(\ket{+}+i\ket{-})/\sqrt{2}$:
        unitary evolution (blue solid),
        driving with $\kappa(t)=\cos t/10$ (red dot),
        and random walk (black solid).
        Position distribution~$P$ at
        (b)~$t=22$, (c)~$t=32$, and (d)~$t=48$.
            }
\label{fig:varanddist}
\end{figure}
$V$ oscillates periodically between
classical diffusive and quantum ballistic values at various
times~$t$ with an $\eta$-dependent period as anticipated by the above argument.
Variances of the undriven-coin quantum walk and the classical random walk
provide tight upper and lower bounds for the driven-coin time-dependent variance.
Figures~\ref{fig:varanddist}(b-d) display numerical results for the position distribution as
a blend of classical and quantum distributions ($t=22$), nearly
fully quantum ($t=32$) and nearly fully classical ($t=48$).

\begin{figure}
	(a) \includegraphics[width=0.4\columnwidth]{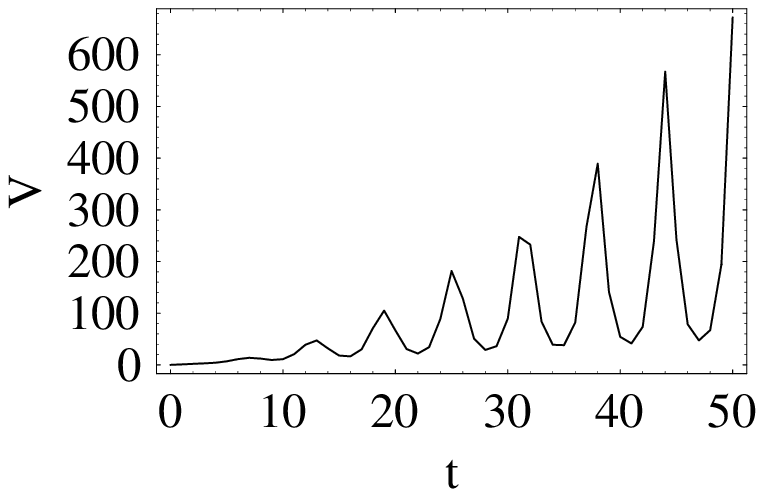}
	(b) \includegraphics[width=0.4\columnwidth]{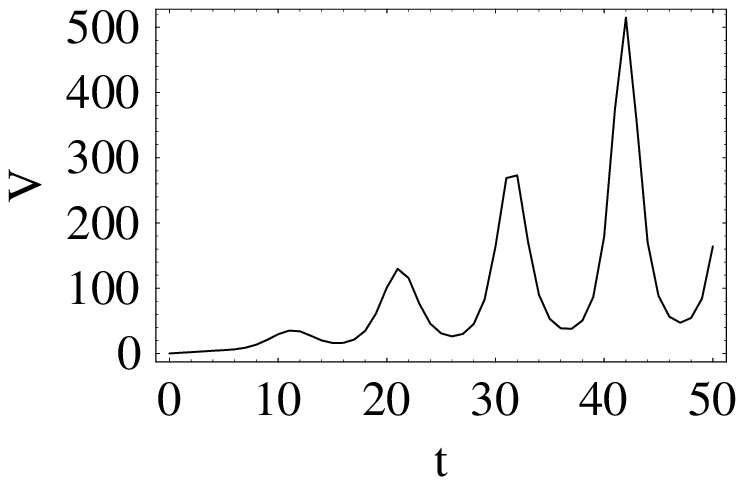}
	\caption{
	(Color online)
	Numerically evaluated variance of the walkerÕs position over time $t$
	with same the same initial state as in Fig,~\ref{fig:varanddist}
	but driving with
	(a)~$\kappa(t) =\cos t/2$ and (b) $\kappa(t)=\cos 3t/10$.
            }
\label{fig:Vhigherfrequency}
\end{figure}

\begin{figure}
	(a) \includegraphics[width=0.4\columnwidth]{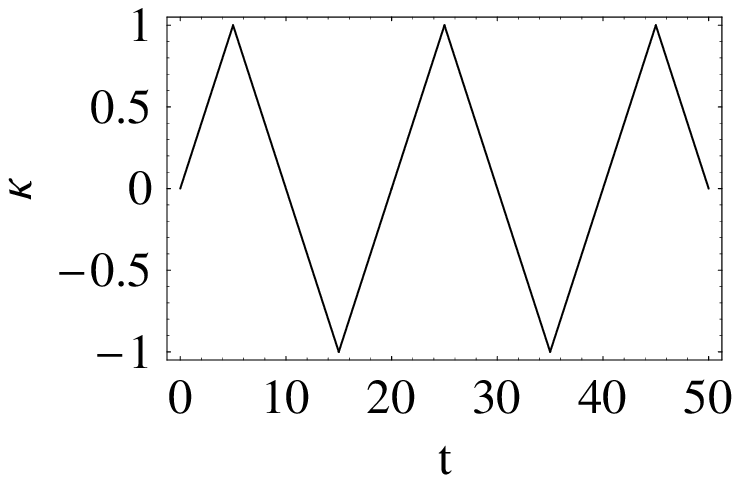}
	(b) \includegraphics[width=0.4\columnwidth]{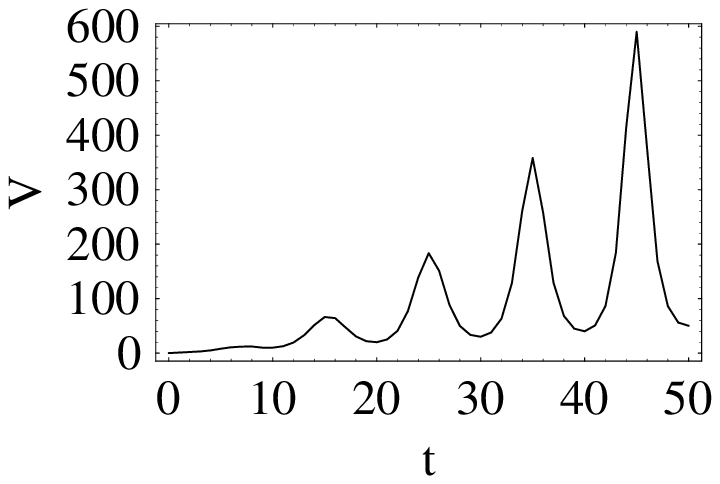}
	\caption{
	(Color online)
	(a)~The sawtooth driving function~$\kappa(t)$ and
	(b)~the corresponding numerically evaluated variance of the walkerÕs position as a function of time~$t$.
	}
\label{fig:sawtooth}
\end{figure}

As Figs.~\ref{fig:varanddist}(a-d) are shown for just one frequency~$\eta$ of the driving field,
we show the generality of our result for a periodic driving field by repeating for
two other higher frequencies.
The corresponding variance functions are shown in Figs.~\ref{fig:Vhigherfrequency}(a,b),
and the periodic oscillation of~$V$ is evident in these plots as well.
For generality we also consider a non-sinusoidal driving function.
Specifically we consider the sawtooth driving function depicted in Fig.~\ref{fig:sawtooth}(a)
and show the corresponding time-dependent variance in Fig.~\ref{fig:sawtooth}(b),
which also shows the periodicity of the variance over time, commensurate with the periodicity observed 
for the sinusoidal driving function.

The decreasing variance at times can seem counter-intuitive.
The variance is often closely associated with a distribution's entropy,
especially for the normal distribution.
As our control is an incoherent measurement process,
the decreasing variance is startling unless the concepts of variance and entropy are dissociated.
We now show that the entropy of the walker's position distribution is non-decreasing despite the
variance decreasing.

The randomness in the system is reflected in the position of the walker.
\begin{figure}
    \includegraphics[width=\columnwidth]{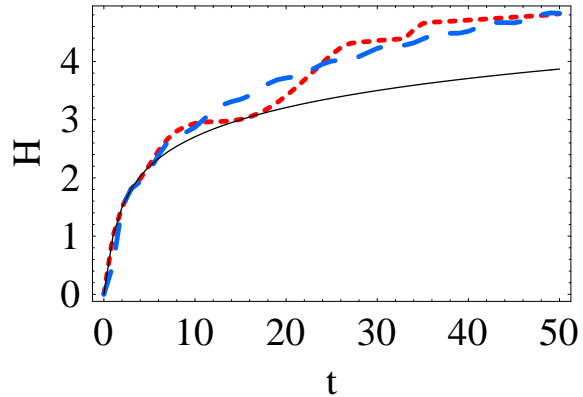}
\caption{
    (Color online) Numerically evaluated entropy $H$ of the walker over time~$t$
        with initial state $\ket{0}(\ket{+}+i\ket{-})/\sqrt{2}$:
        unitary evolution (blue dash),
        driving with $\kappa(t)=\cos(t/10)$ (red dot),
        and random walk (black solid).
    }
    \label{fig:entropy}
\end{figure}
We employ the entropy
\begin{equation}
\label{eq:entropy}
    H(t)=-\sum_{x} P_\text{w}(x,t) \text{log}_2 P_\text{w}(x,t)
\end{equation}
to characterize the randomness of
the walker system. We plot the entropies for a classical RW, QW with
unitary evolution and QW with a driven coin in
Fig.~\ref{fig:entropy}. From the numerical simulations we can see
that, for these three kinds of walks, the entropies which represent
randomness are increasing with time.
Though the periodically driven
coin measurement causes the variance to increase and to decrease alternately over time,
the randomness of the walker's position position distribution,
quantified by entropy~(\ref{eq:entropy}), never decreases.

\section{Asymptotic analysis with the `momentum' representation}

Using `momentum' states in the Fourier domain
\begin{equation}
    \ket{k}=\sum_{x=-\infty}^\infty \text{e}^{ikx}\ket{x}
\end{equation}
such that
\begin{equation}
    S\ket{k}=\text{e}^{-ik}\ket{k},\;
    k\in\mathbb{R},
\end{equation}
in whose basis the evolution is given by
\begin{equation}
    U\ket{k}_\text{w}\otimes\ket{\Phi}_\text{c}
        =\ket{k}\otimes U(k)\ket{\Phi}_\text{c}
\end{equation}
for
\begin{equation}
    U(k)=\frac{1}{\sqrt{2}}
    \begin{pmatrix}
        \text{e}^{-ik} & \text{e}^{-ik}\\\text{e}^{ik} & -\text{e}^{ik}
    \end{pmatrix}.
\end{equation}
If the initial walker-coin state corresponds to the walker localized at the origin $x=0$
and the coin in any pure state~$|\Phi\rangle_\text{c}$,
the  inverse transform expression
\begin{equation}
    \ket{x}=\int^\pi_{-\pi}\dslash k\text{e}^{-ikx}\ket{k}
\end{equation}
(for $\dslash k:=\frac{\text{d}k}{2\pi}$)
yields the $t$-dependent joint walker-coin state
\begin{equation}
    \rho(t)=\int_{-\pi}^\pi\dslash k
        \int_{-\pi}^\pi\dslash k'\ket{k}\bra{k'}\otimes
        U^t(k)\rho_c(U^\dagger(k))^t.
        \end{equation}
After~$t$ steps the state is
\begin{equation}
\rho_\text{wc}\left(t\right)
        =\int\dslash k\int\dslash k'
            \ket{k}\bra{k'}\otimes\mathcal{L}^t(k,k')\rho_\text{c}
\end{equation}
for
\begin{equation}
    \mathcal{L}(k,k')\hat{O}
         :=\sum_{i\in\{0,\pm\}}U(k) A_{i}\hat{O}A^\dagger_{i}U^\dagger(k').
\end{equation}
For $k=k'$ this superoperator satisfies
\begin{equation}
    \text{Tr}[\mathcal{L}^t(k,k)\hat{O}]=\text{Tr}[\hat{O}]
\end{equation}
for
any~$\hat{O}$ so $\mathcal{L}(k,k)$ is trace-preserving.

Figure~\ref{fig:varanddist} is derived numerically hence not conclusive in revealing
asymptotic spreading behavior arising from the periodically driven coin.
Therefore, we exploit the momentum representation to derive the
asymptotic long-time position distribution
\begin{equation}
    P_\text{w}(x;t)
        =\int_{-\pi}^\pi\dslash k\int_{-\pi}^\pi
            \dslash k\text{e}^{-i(k-k')x}\text{Tr}\left[\mathcal{L}^t(k,k')\rho_c\right]
\end{equation}
of the driven-coin quantum walker.
Using
\begin{equation}
    \sum_{x=-\infty}^\infty x^m\exp[-ix(k-k')]=2\pi(-i)^m\delta^{(m)}(k-k'),
\end{equation}
we obtain
\begin{equation*}
    \left\langle \hat{x}^m\right\rangle=(-1)^m\int_{-\pi}^\pi\dslash k
        \int_{-\pi}^\pi\dslash k'\delta^{(m)}(k-k')\text{Tr}\{\mathcal{L}^t(k,k')\rho_c\}
\end{equation*}
so the first two moments are
\begin{equation}
\label{eq:first}
    \left\langle \hat{x}\right\rangle
        =-\int_{-\pi}^{\pi}\dslash k\sum_{j=1}^t\text{Tr}\{Z\mathcal{L}^j(k,k)\rho_c\}.
\end{equation}
and
\begin{align}
\label{eq:second}
    \left\langle \hat{x}^2\right\rangle
        =&-\int_{-\pi}^{\pi}\dslash k\Big(
    \sum_{j=1}^t\sum_{j'=1}^j
    \text{Tr}\left\{Z\mathcal{L}^{j-j'}(k,k)Z\mathcal{L}^{j'}(k,k)\rho_c\right\}
            \nonumber\\
        &+\sum_{j=1}^t\sum_{j'=1}^{j-1}
        \text{Tr}\left\{Z\mathcal{L}^{j-j'}(k,k)\mathcal{L}^{j'}(k,k)\rho_cZ\right\}\Big)
\end{align}
with $Z=\begin{pmatrix}1&0\\0 &-1\end{pmatrix}$.

As~$\mathcal{L}(k,k)$ is additive, we obtain
\begin{align}
\label{eq:rho}
    \mathcal{L}(k,k) \mathcal{E}\rho_\text{c}=&\kappa(t)\mathcal{L}(k,k)\rho_\text{c}
            \nonumber\\
        &+[1-\kappa(t)]\mathcal{L}(k,k)\sum_{\epsilon\in\pm}\mathcal{P}^\epsilon
            \rho_\text{c}\mathcal{P}^\epsilon
\end{align}
with the first term on the right-hand side corresponding to a
quantum-walk mapping (these terms are indicated by superscript~Q)
for a time-dependent coin and the second term
corresponding to coherence-destroying measurements that transform
the quantum walk into the random walk
(these terms are indicated by a superscript R).
Now we exploit linearity to find
asymptotic solutions to the first and second term separately.

To study the first term on the right-hand side of~(\ref{eq:rho}),
we first specialize to the unitary case
\begin{equation}
    \mathcal{L}(k,k)\hat{O}=U(k)\hat{O} U^\dagger(k)
\end{equation}
and
express
\begin{equation}
    \ket{\Phi}_\text{c}=\sum_{l=0}^1c_l(k)\ket{\phi_l(k)}.
\end{equation}
Here $\{\ket{\phi_l(k)}\}$ is the orthonormal eigenbasis for $U(k)$
with eigenvalues $\{\text{e}^{i\theta_l(k)}\}$
such that
\begin{equation}
    \theta_0(k)+\theta_1(k)=\pi.
\end{equation}

For the standard quantum walk with a Hadamard coin,
\begin{align}
\label{eq:eigenstate}
\ket{\phi_l(k)}
    =&\sqrt{1+\cos^2k-(-1)^l\cos k\sqrt{1+\cos^2k}}
            \nonumber\\
        &\times\begin{pmatrix}\text{e}^{-ik}/\sqrt{2}\\
            \text{e}^{-i\theta_l(k)}-\text{e}^{-ik}/\sqrt{2}\end{pmatrix}.
\end{align}
Defining
\begin{equation}
    c_{ll'}(k):=c^*_l(k)c_{l'}(k)
\end{equation}
and
\begin{equation}
    \theta_{l'l}(k):=\theta_{l'}(k)-\theta_l(k),
\end{equation}
the coin state
\begin{equation}
    \rho_\text{c}(t)=\sum_{ll'}c_{ll'}(k)\ket{\phi_{l'}(k)}\bra{\phi_l(k)}\text{e}^{i\theta_{l'l}(k)t}
\end{equation}
after~$t$ steps is substituted into Eqs.~(\ref{eq:first}) and~(\ref{eq:second}) to yield
\begin{equation}
   \left \langle \hat{x}\right\rangle^\text{Q}
        =t-2\int_{-\pi}^\pi\dslash k\sum_{l,l'}c_{ll'}(k)\mathcal{P}^+_{ll'}(k)
    \sum_{j=1}^t\text{e}^{ij\theta_{l'l}(k)}
\end{equation}
with
$\mathcal{P}^+_{ll'}(k)=\bra{\phi_l(k)}\mathcal{P}^+\ket{\phi_{l'}(k)}$
and
\begin{align}
    \left\langle \hat{x}^2\right\rangle^\text{Q} =&\int_{-\pi}^\pi\dslash k
        \sum_{l,l',l''}c_{ll'}(k)Z_{ll''}(k)Z_{l''l'}(k)\nonumber\\
    &\times \sum_{j,j'=1}^t\text{e}^{i(j-j')'\theta_{l'l}(k)}
\end{align}
with
$Z_{ll'}(k)=\bra{\phi_l(k)}Z\ket{\phi_{l'}(k)}$.

As $U(k)$ is nondegenerate so $\theta_{0,1}(k)$ are distinct, most
terms for $\left\langle \hat{x}\right\rangle^\text{Q}$ and
$\left\langle \hat{x}^2\right\rangle^\text{Q}$ above oscillate
strongly, and only diagonal terms survive in the long-time limit.
For $\langle \hat{x}\rangle^\text{Q}$, the condition to be on the
diagonal should be $l=l'$. Whereas for $\langle
\hat{x}^2\rangle^\text{Q}$ there are two sets of non-oscillatory
terms: terms with $l=l'=l''$ (for quadratic dependence on $t$) and
terms with $j=j'$ and $l=l'$ (for linear dependence on $t$).

Inserting Eq.~({\ref{eq:eigenstate}}) into the equations above
yields
\begin{equation}
\text{V}^\text{Q}\longrightarrow_{t\rightarrow\infty} t^2(C_2-C_1^2)
\end{equation} and
\begin{widetext}
\begin{align}
    C_1 =&1-2\int^{\pi}_{-\pi}\dslash k\sum_{l=1,2}|c_l(k)|^2P^+_{ll}(k)
        =1-\int_{-\pi}^{\pi}\frac{\dslash k}{1+\cos^2k}
        =1-\frac{1}{\sqrt{2}},
                    \nonumber\\
    C_2=&1-4\int^{\pi}_{-\pi}\dslash k\sum_{l=1,2}|c_l(k)|^2P^+_{ll}(k)P^-_{ll}(k)
        =1-\int_{-\pi}^{\pi}\frac{\dslash k}{1+\cos^2k}=1-\frac{1}{\sqrt{2}}.
\end{align}
By path integration~\cite{NV00},
\begin{align*}
    P^\text{Q}_\text{w}(x,t)
        \approx&\int_{-\pi}^{\pi}
    \dslash k\int_{-\pi}^\pi\dslash k'\text{e}^{-ix(k-k')}
    \sum_{l,l'}c^*_l(k)c_{l'}(k')
        \text{e}^{-i\left[\theta_{l'}(k')-\theta_l(k)\right]t}\bra{\phi_l(k)}\phi_{l'}(k')\rangle\\
    \approx&\int_{-\pi}^{\pi}\text{d}k\frac{1
            +(-1)^{x+t}}{\pi t\frac{\sin k}{(1+\cos^2k)^{3/2}}}
            \Bigg\{\left(1-\frac{x}{t}\right)^2
    \cos^2\Bigg[\arcsin\left(\frac{\sin k}{\sqrt{2}}\right)t+xk-\frac{\pi}{4}\Bigg]
        \nonumber\\
    &+\left[1-\left(\frac{x}{t}\right)^2\right]
        \cos^2\Bigg[\arcsin
            \left(\frac{\sin k}{\sqrt{2}}\right)t
    +(x-1)k-\frac{\pi}{4}\Bigg]\Bigg\},
\end{align*}
and we now have the solution for the quantum part of the walk.
\end{widetext}

Now we proceed to study the second (random-walk) term of Eq.~(\ref{eq:rho}).
Using the representation
\begin{equation}
    \hat{O}=r_1\one+r_2X+r_3Y+r_4 Z
\end{equation}
(Pauli representation)~\cite{BCA03a},
we obtain
\begin{equation}
\mathcal{L}(k,k)\hat{O}
        =\begin{pmatrix}
           1 & 0 & 0 & 0\\
           0 & 0 & 0 & \cos2k\\
           0 & 0 & 0 & -\sin2k\\
           0 & 0 & 0 & 0
        \end{pmatrix}\begin{pmatrix}
           r_1 \\r_2\\ r_3 \\r_4
        \end{pmatrix},
\end{equation}
which leads to new expressions for the first two moments
\begin{align}
   \left \langle \hat{x}\right\rangle^\text{R}
    =&-\int_{-\pi}^{\pi}\dslash k
    \begin{pmatrix}0&0&0&1\end{pmatrix}
    \left(\sum_{j=1}^t
    \mathcal{L}^j(k,k)\right)
    \begin{pmatrix}
        r_1\\r_2\\r_3\\r_4
    \end{pmatrix}
        \nonumber\\
    =&0
 \end{align}
 and
\begin{align}
    \left \langle \hat{x}^2\right\rangle^\text{R}=&t-\int_{-\pi}^{\pi}\dslash k
    \begin{pmatrix}
        1&0&0&0
    \end{pmatrix}
    \Bigg[Z_L\sum_{j=1}^t\sum_{j'=1}^{j-1}\mathcal{L}^{j-j'}(k,k)
            \nonumber\\
        &\times\left(Z_\text{L}+Z_\text{R}\right)\mathcal{L}^{j'}(k,k)\Bigg]
    \begin{pmatrix}r_1\\r_2\\r_3\\r_4\end{pmatrix}=t
\end{align}
for
\begin{equation}
    Z_\text{L}=\begin{pmatrix}
        0&0&0&1\\0&0&i&0\\0&-i&0&0\\1&0&0&0
        \end{pmatrix},
    Z_\text{R}=\begin{pmatrix}
        0&0&0&1\\0&0&-i&0\\0&i&0&0\\1&0&0&0
        \end{pmatrix}.
\end{equation}
As~$\left\langle \hat{x}\right\rangle^\text{R}=0$, $\left\langle
\hat{x}^2\right\rangle^\text{R}$ equals the variance and is purely
diffusive due to its proportionally with~$t$.

\begin{figure}
    \includegraphics[width=0.4\columnwidth]{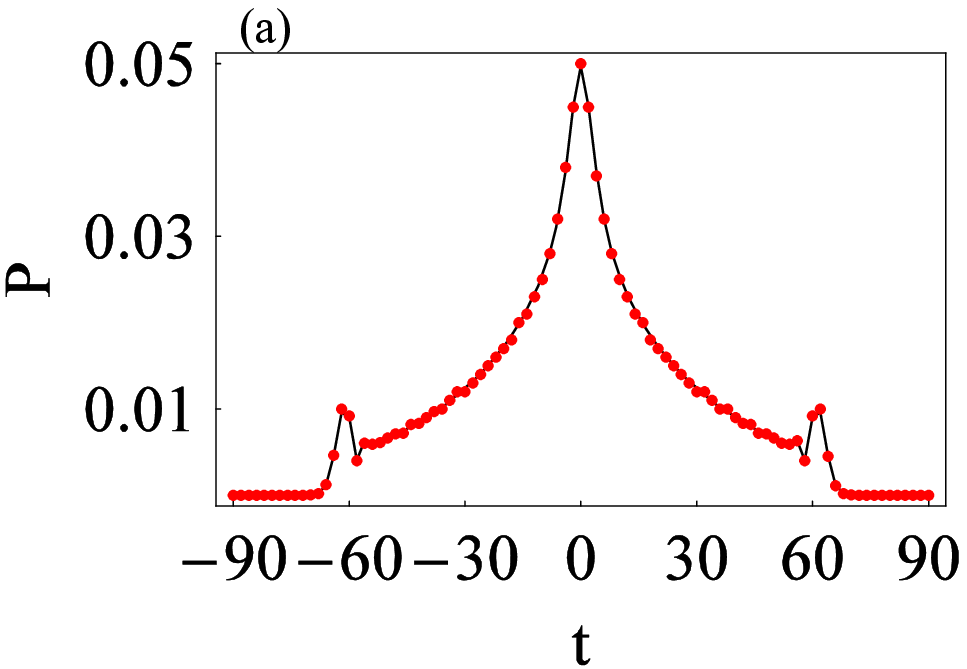}
    \includegraphics[width=0.4\columnwidth]{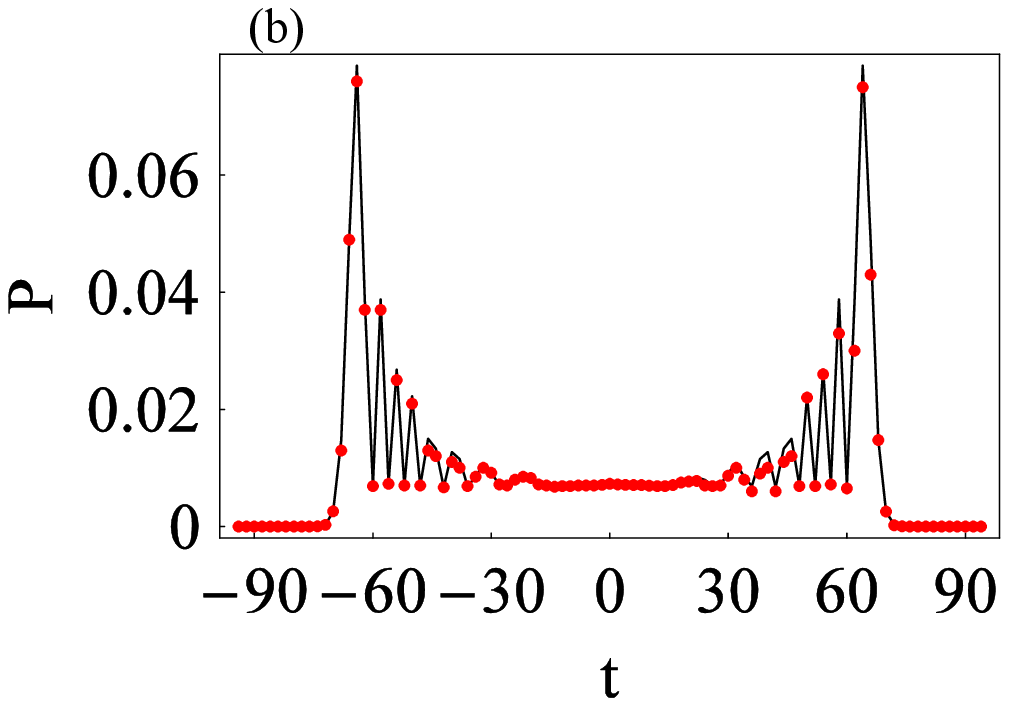}
    \caption{
    (Color online)
    Walker position distributions~$P_\text{w}(x,t)$
    for even values of~$x$ and (a) $t=90$ and (b) $t=94$
    with initial state $\ket{0}(\ket{+}+i\ket{-})/\sqrt{2}$ and
    driven-coin function $\kappa(t)=\cos(t/10)$ for precise numerical simulations (dots)
    and asymptotic expression (solid).
    }
\label{fig:anaposition}
\end{figure}

The asymptotic binomial random-walk position distribution evaluated only
over even (odd)~$x$ for even (odd)~$t$ is thus
\begin{equation}
    P^\text{R}_\text{w}(x,t)
        =\frac{1}{2^t}\frac{t!}{\left[\frac{t-x}{2}\right]!\left[\frac{t+x}{2}\right]!}.
\end{equation}
Thus, for a QW with a driven coin, we have the asymptotic position distribution
\begin{equation}
 P_\text{w}(x,t)=\kappa P^\text{Q}_\text{w}(x,t)+(1-\kappa)P^\text{R}_\text{w}(x,t)
\end{equation}
and variance
\begin{equation}
\text{V}=\kappa
\text{V}^\text{Q}+(1-\kappa) t.
\end{equation} Excellent agreement
between analytical asymptotic expressions and numerical results in
the long-time limit is obtained, e.g.\ for the walker's position
distribution (Figure~\ref{fig:anaposition}).

\section{Entanglement between the walker and the coin}

Although the usual signature of the quantum walk is the rate of
spreading~$V$, which, for the driven coin, is a weighted
sum of the quantum ballistic spread~$V^\text{Q}$ and the diffusive
term $V^\text{R}$ proportional to~$t$, the quantum-classical divide
can be explored in more depth through studying entanglement between
the walker and the coin. Entanglement should be zero in the
random-walk case and should generally be non-zero in the
quantum-walk case.
We analyze walker-coin quantum correlation with two measures: measurement induced
disturbance (MID)~\cite{Luo08a} and quantum discord
(QD)~\cite{OZ01}.

Discord uses von Neumann measurements to quantify QD~\cite{OZ01},
which consist of one-dimensional projectors summing to the identity
operator. We use projection operators $\{B_j\}$ to denote a von
Neumann measurements for coin state only. Quantum conditional
entropy is $S[\rho_\text{wc}(t)|\{B_j\}]:=\sum_j p_j S[\rho_j(t)]$
with $S(\cdot)$ the von Neumann entropy, and the associated quantum
mutual information is
\begin{equation}
\mathcal{I} [\rho_\text{wc}(t)|\{B_j\}]:=S [\rho_\text{w}(t)]-S
[\rho_\text{wc}(t)|\{B_j\}],
\end{equation} where the
conditional density operator operator $\rho_j(t)=(\one\otimes
B_j)\rho_\text{wc}(t)(\one\otimes B_j)$ with the measurement result
$j$, and the probability $p_j=\text{Tr}[(\one\otimes
B_j)\rho_\text{wc}(t)(\one\otimes B_j)]$.

For classical correlations
\begin{equation}
\mathcal{C}_\text{cl}[\rho_\text{wc}(t)]:=\text{sup}_{B_j}\mathcal{I}[\rho_\text{wc}(t)|\{B_j\}],
\end{equation}
QD is
\begin{equation}
D:=\mathcal{I}[\rho_\text{wc}(t)]-\mathcal{C}_\text{cl}[\rho_\text{wc}(t)].
\end{equation} With respect to QD, correlations between $\rho_\text{w}$ and
$\rho_\text{c}$ are classical if there exists a unique local
measurement strategy on the coin
 $\{B_j\}$ leaving
$\rho_i(t)$ unaltered from the original joint walker-coin
state~$\rho_\text{wc}(t)$. Calculating QD is difficult due the need
to maximize over all possible von Neumann-type measurements of the
coin state in order to determine the classical correlation.

\begin{figure}
    \includegraphics[width=0.4\columnwidth]{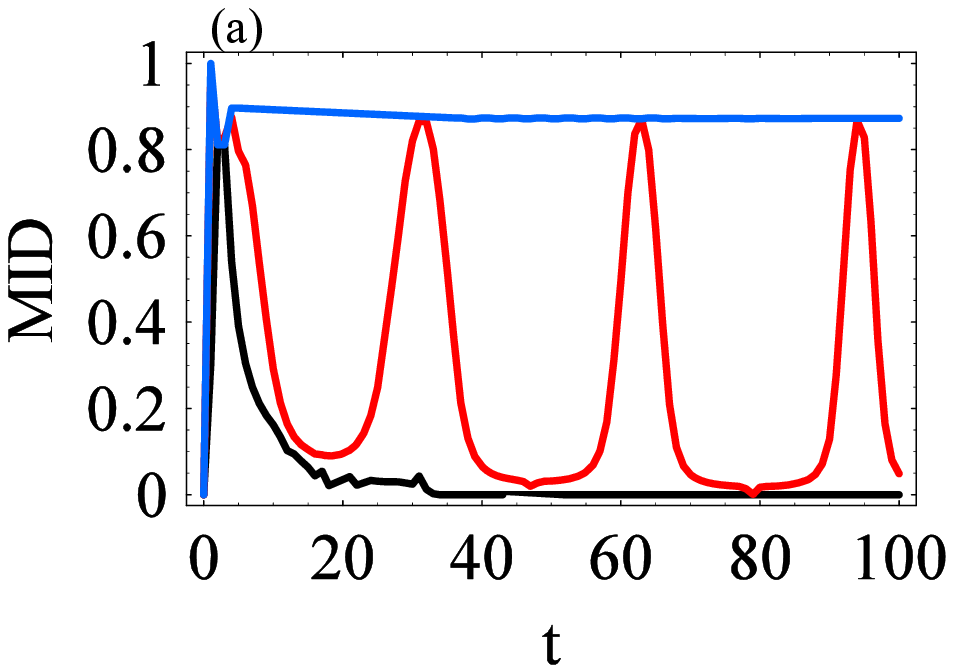}
    \includegraphics[width=0.4\columnwidth]{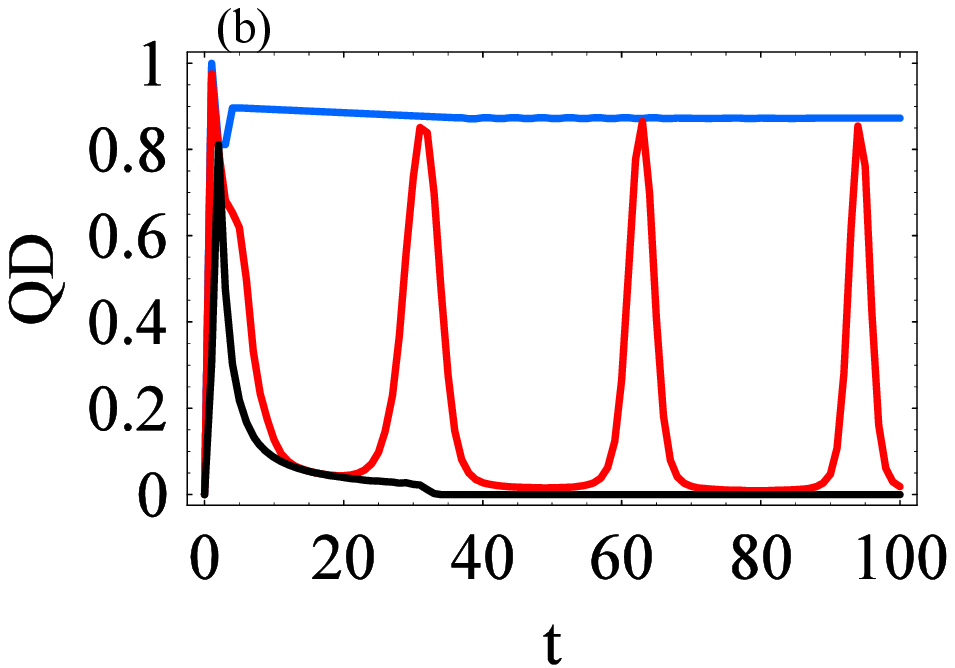}
    \caption{
        (Color online)
        (a)~The MID and
        (b)~QD for a QW on a line after~$t=100$ steps for
        undriven (blue), driven with $\kappa(t)=\cos(t/10)$ (red) and random walk (black) cases
    with initial state
    $\ket{0}(\ket{+}+i\ket{-})/\sqrt{2}$.
    }
\label{fig:MID}
\end{figure}

MID~\cite{Luo08a}
\begin{equation}
Q[\rho_\text{wc}(t)]:=\mathcal{I}[\rho_\text{wc}(t)]-\mathcal{I}[\rho_\text{wc}(t)|\{B_j\}]
\end{equation} has an advantage over QD in that MID is
operational but tends to overestimate non-classicality because of a
lack of optimization over local measurements. We therefore ascertain
whether the joint walker-coin state is `quantum' by determining if a
local measurement strategy exists that leaves the state unchanged.
We numerically evaluate MID and QD for a QW in the undriven,
driven-coin and random-walk cases and display results in
Figure~\ref{fig:MID}.
For the undriven-coin case, MID and QD are the
same. For the driven-coin case, MID and QD exhibit
quantum-correlation oscillations with the same period as for the
position-variance oscillation, i.e.\ the oscillation between
diffusive and ballistic spreading.

\section{Conclusion}

In summary we study the driven-coin quantum walk and
discover that controlling the quantum coin causes the walker's spread to oscillate
between diffusive and ballistic spreading, which are standard signatures of
classical random vs.\ quantum walks, respectively.
Our results are determined by numerical means for all times and by closed-form
expressions in the asymptotically long-time limit.
We prove that the walker's reduced position distribution is an incoherent
mixture of classical- and quantum-walk distributions.

This oscillation between classical-diffusive and quantum-ballistic spreading
is manifest in the quantum correlations between the walker and the coin.
One can understand the surprising result of alternating increasing-decreasing variance
as being due to a periodic restoration of a highly non-Gaussian walker position distribution,
due to quantum walking, into a Gaussian distribution,
with a concomitant decrease of variance.
We show, however, that this decrease of variance is not accompanied by a decrease in entropy:
entropy is strictly non-decreasing as expected for an unconditional measurement-based control.
Our alternating increasing-decreasing walker-position variance results
could be valuable for tuning quantum walks~\cite{Ken07}
and especially challenge common notions of decoherence in quantum walks associated with tying
variance to decoherence.

\acknowledgements PX acknowledges financial support from NSFC under
No. 11004029 and No. 11174052 and China's 973 Program under No.
2011CB921203, and BCS acknowledges financial support from CIFAR,
NSERC and AITF.

\bibliography{qw}
\end{document}